\documentclass[fleqn,twoside]{article}
\usepackage{amsmath}
\usepackage{espcrc2}

\usepackage{graphicx}

\newcommand{\be}{\begin{equation}}
\newcommand{\ee}{\end{equation}}
\newcommand{\bea}{\begin{eqnarray}}
\newcommand{\eea}{\end{eqnarray}}

\newcommand{\etap}{$\eta'$ }

\newcommand{\tr}{\mathbin{\text{Tr}}}

\def\slashchar#1{\setbox0=\hbox{$#1$}           
   \dimen0=\wd0                                 
   \setbox1=\hbox{/} \dimen1=\wd1               
   \ifdim\dimen0>\dimen1                        
      \rlap{\hbox to \dimen0{\hfil/\hfil}}      
      #1                                        
   \else                                        
      \rlap{\hbox to \dimen1{\hfil$#1$\hfil}}   
      /                                         
   \fi}

\newcommand{\slashD}{\slashchar{D}}

\title{Index Theorem and Random Matrix Theory for Improved Staggered Quarks}

\author{E. Follana \address[UG]{Department of Physics and Astronomy,
University of Glasgow, \\G12 8QQ Glasgow, UK}%
  \thanks{The results presented here have been obtained in collaboration with
    A. Hart and C. Davies.}}

\begin{document}

\begin{abstract}
 We study various improved staggered quark Dirac operators on quenched
  gluon backgrounds in lattice QCD generated using a Symanzik-improved
  gluon action. We find a clear separation of the spectrum of
  eigenvalues into would-be zero modes and others. The number of
  would-be zero modes depends on the topological charge as expected
  from the Index Theorem, and their chirality expectation value is
  large. The remaining modes have low chirality and show clear signs
  of clustering into quartets and approaching the random matrix theory
  predictions for all topological charge sectors. We conclude that
  improvement of the fermionic and gauge actions moves the staggered
  quarks closer to the continuum limit where they respond correctly to
  QCD topology.

\vspace{1pc}
\end{abstract}

\maketitle

\section{INTRODUCTION}

Topology plays a key role in our understanding of some important
features of QCD, such as the axial anomaly, or the \etap mass.

In the continuum, the Index Theorem relates the topological charge of
a smooth gauge field with the number of chiral zero modes of the
corresponding Dirac operator.

For QCD in the $\epsilon$ regime, there are detailed predictions of
the distribution of the low-lying eigenvalues of the Dirac operator,
in each sector of fixed topological charge.

A correct discretization of QCD must reproduce these features, at
least in the continuum limit. Furthermore, in order to use such
discretizations in practice to tackle calculations related to
topology, we would like to see such topological properties manifesting
themselves at values of the parameters at which we can realistically
do simulations.

It has been widely held that the staggered discretization of QCD is
insensitive to topology. Previous studies did not find, in the raw,
non-smoothed gauge field configurations, signs of an Index Theorem,
and the predictions of RMT were not reproduced; the calculations in
all sectors of fixed topological charge gave the same results, in
agreement with the theoretical predictions for the sector of zero
topological charge.

There are large-scale simulations using staggered fermions today
\cite{MILC,Davies:2003ik}, and more are planned for the near
future. It is therefore important to understand to what extent
staggered quarks show the correct topological properties.

\section{STAGGERED DIRAC OPERATORS}
All the Dirac operators we study have the general form:
\be
  S = \sum_{x,y} \bar{\chi}(x) \slashD(x,y) \chi(y)
\ee
with $\slashD$ a gauge-invariant linear operator. The first example is
the one-link staggered Dirac operator (also called naive, unimproved
staggered or Kogut-Susskind in the literature):
\bea &\hspace{-1mm} \slashD(x,y) = \frac{1}{2 u_0} \sum_{\mu}
\hspace{-1mm} \alpha_\mu(x)
\hspace{-1mm} \left( U_\mu(x) \delta_{x+\mu,y} - H.c. \right) \\
&\alpha_\mu(x) = (-1)^{\sum_{\nu > \mu} x_\nu} \eea
This Dirac operator has some simple properties, which are shared with
all the improved operators we will introduce later. First, it is
antihermitian, $\slashD^\dagger = -\slashD$.  It also obeys a remnant
of the continuum $\gamma_5$ anticommutation relation:
\be
\left\{\slashD,\epsilon \right\}  = 0, \;\;\;  
{\rm with} \;\;\;
\epsilon(x)  = (-1) ^ {\sum_\mu x_\mu}
\ee

As a consequence of these two properties, its spectrum is purely
imaginary and eigenvalues come in complex-conjugate pairs,
\be 
sp(\slashD) =
\{\pm i \lambda, \lambda \in \Re \} 
\label{pairs}
\ee
  
This operator corresponds (in four dimensions) to 4 ``tastes'' of
fermions. There are unphysical taste-changing interactions, involving
at leading order the exchange of a gluon of momentum
$q \approx\pi/a$. 
Such interactions are perturbative for typical values of the lattice
spacing, and can be corrected systematically \`a la Symanzik. This can
be accomplished by smearing the gauge field to remove the coupling
between quarks and gluons with momentum $\pi / a$
\cite{Naik,Lepage:1998vj,Blum:1997uf,Bernard:1999xx,Orginos:1999cr}.
Including appropriate paths up to length seven leads to the so-called
{\textsc{fat7}} operator. By adding two more terms (the five-link
Lepage term and the three-link Naik term), we obtain an operator
improved to order $a^2$ at tree level, called {\textsc{asq}}. If we
also add tadpole improvement we get the {\textsc{asqtad}} operator.

\begin{figure}
  \includegraphics*[width=2.9in]{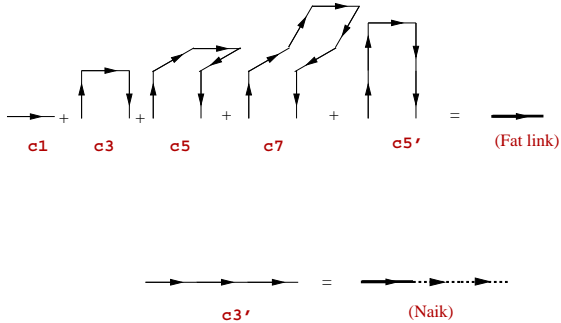}
  \caption{paths entering the {\textsc{asq(tad)}} operator}
\end{figure}

Another improved staggered Dirac operator, motivated by perfect action
ideas, is the {\textsc{hyp}} operator, which involves three levels of
(restricted) APE smearing with projection onto $SU(3)$ at each
level. The restrictions are such that each fat link includes
contributions only from thin links belonging to hypercubes attached to
the original link. \cite{Knechtli:2000ku}.

The final improved staggered operator we will consider here is the so
called {\textsc{hisq}} (Highly Improved Staggered Quarks) operator, which
involves two levels of smearing: first a FAT7 smearing on the original
links, followed by a projection onto $SU(3)$, then {\textsc{asq}} on these
fattened links \cite{Follana:2003,Follana:2004}.

Both {\textsc{hyp}} and {\textsc{hisq}} show much smaller
taste-changing effects than {\textsc{asqtad}}.

\section{DETAILS OF THE SIMULATION}

Most of our results come from gauge configurations generated using a
gauge action which is Symanzik-improved at tree level, including
tadpole improvement. We have three different ensembles of about 1000
configurations each, whose parameters are shown in
table~\ref{table:1}. We will refer to the configurations by their
volume, and whether they correspond to the fine, $a = 0.093 $ fm
ensemble or to the coarse, $0.125 $ fm one.
\begin{table}[t]
\caption{Parameters of the improved gauge ensembles}
\label{table:1}
\newcommand{\m}{\hphantom{$-$}}
\begin{tabular}{@{}llll}
\hline
Volume            &  \m$12^4$  & \m$16^4$    & \m$12^4$ \\
$a$ (fm)           & \m0.093 & \m0.093 & \m0.125 \\
Length (fm)       & \m1.12 & \m1.49 & \m1.50 \\
\hline
\end{tabular}\\[2.2pt]
\end{table}
The topological charge $Q$ is determined by cooling the fields and
then using a highly accurate discretization of the continuum
expression. We do the cooling with two different actions to check for
consistency. The value of $Q$ for lattice gauge fields is not, in
general, unambiguously defined, and therefore the values obtained with
the two methods do not always coincide. This happens in about $10\%$
of the configurations.

We have also used a few Wilson (unimproved) gauge configurations, with
a lattice spacing of $a~=~0.093$, and a volume of $16^3 \times 32$.

\section{SPECTRUM AND INDEX THEOREM}

\subsection{Continuum}

The eigenmodes of the anti-hermitian, gauge-covariant, massless
continuum Dirac operator are given by
\be
\slashD f_s = i \lambda_s f_s \; , ~~~~~ \lambda_s \in R \; .
\ee
where we use orthonormalised eigenvectors, 
$f_s^\dagger f_t = \delta_{s,t}$.
As 
$\{ \slashD,\gamma_5 \} = 0$
, the spectrum is symmetric about zero: if $\lambda_s \not = 0$, then
$\gamma_5 f_s$ is also an eigenvector with eigenvalue $-i \lambda_s$,
and chirality $\chi_s \equiv f_s^\dagger \gamma_5 f_s = 0$. The zero
modes, $\lambda_s = 0$, can be chosen with definite chirality: $\chi_s
= \pm 1$. In general there are $n^{\pm}$ such modes, whose relative
number is fixed by the (gluonic) topological charge
\begin{equation}
Q = \frac{1}{32 \pi^2} \int d^4x \; \epsilon_{\mu \nu \sigma \tau}
\tr F_{\mu \nu}(x) F_{\sigma \tau}(x)
\end{equation}
via the Atiyah--Singer Index Theorem
\cite{Atiyah:1963,Atiyah:1968}
\begin{equation}
Q = m \tr \frac{\gamma_5}{\slashD + m} = n^+ - n^- \; ,
\label{eqn_index}
\end{equation}
where $m$ is the quark mass. 

\subsection{Chiral lattice discretizations}

On the lattice, Dirac operators that satisfy the Ginsparg-Wilson
relation
\be
\left\{\gamma_5, \slashD \right\} = \overline{a} \slashD \gamma_5 \slashD
\ee
can have exact, chiral zero modes, which then may be used to define a
topological charge \cite{FP}:
\bea
Q & = a^4 \sum_x q(x) =  n^+ - n^- \\
q(x) & = - \frac{1}{2} \; \overline{a} \;\tr \left\{\gamma_5 \slashD(x,x) \right\}
\eea
where $q(x)$ is a local, gauge-invariant function of the gauge fields.

For the fixed-point Dirac operator, furthermore, there is a genuine
Index Theorem at finite cutoff \cite{FP},
\be
Q^{FP} = n^+ - n^-
\ee
where $Q_{FP}$ is a purely gluonic operator with the characteristics
of a proper topological charge.

\subsection{Staggered discretization}

The staggered Dirac operator has no exact zero modes, and therefore we
cannot expect to have an exact Index Theorem. However, close to the
continuum limit, we should see an approximate version of the continuum
behaviour: the first few eigenmodes with high chirality, in the number
required by the continuum Index Theorem, and the rest of the
eigenmodes with small chirality.

This is known to happen with sufficiently smooth gauge fields ,
obtained, for example, by repeated smearing
\cite{Damgaard:2000ah,Damgaard:2001ep}, or by a lattice discretization
of continuum instantons \cite{Smit:1987fn}.

However, we are interested here in the properties of the raw,
non-smoothed configurations, where such behaviour was not found in
previous studies with unimproved staggered quarks
\cite{Hands:1990,Venkataraman:1998yj,Hasenfratz:2003}. We want to
understand to what degree the continuum features are already present
in gauge fields from ensembles generated with parameters used in
typical present-day simulations.

To test this we compute the topological charge and the chirality of
the first few eigenmodes for the gauge fields in our ensembles. The
staggered version of the continuum $\gamma_5$ operator relevant for
this calculation must be a taste-singlet operator, in order for it to
couple to the vacuum. We use a gauge-invariant, point-split four link
operator \cite{Golterman:1984}.

Taking into account that the staggered discretization describes 4
tastes, we expect to have a quadruple near-degeneracy in the
spectrum. Furthermore, if an approximate Index Theorem applies, we
expect to see $4~n^+$ approximate zero modes with chirality near 1,
and $4~n^-$ approximate zero modes with chirality near -1, with $n^+$,
$n^-$ such that
\be
Q = \frac{1}{4} (n^+ - n^-)
\ee

\begin{figure}
\includegraphics[width=2.in,clip, angle=270]{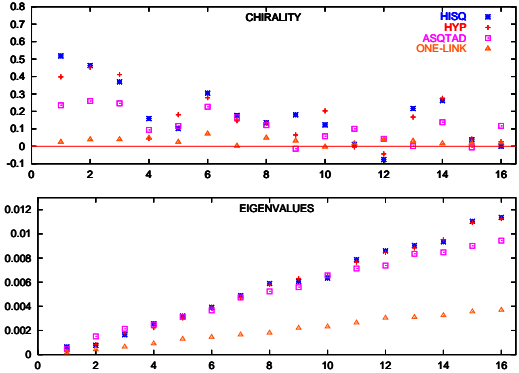}

\caption{The absolute value of a typical low-lying eigenmode (half )
  spectrum for a $16^3 \times 32$ Wilson gauge configuration of $Q =
  2$, for various staggered fermion formulations. The bottom panel
  gives the absolute value of the eigenvalue, $\lambda_s$, ordered
  according to increasing size. The $x$ axis is then simply eigenvalue
  number. The top panel is the chirality of the modes.}

\label{chirality_wilson}
\end{figure}

\begin{figure}
\includegraphics*[width=2in,angle=270]{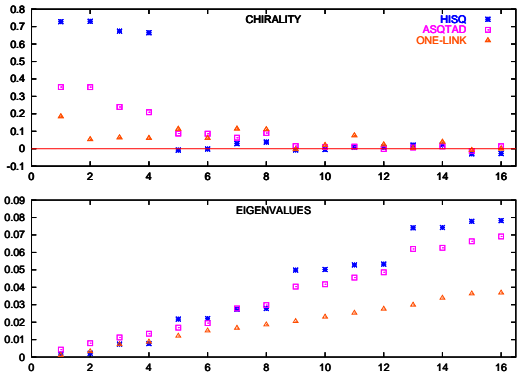}

\caption{as in fig. \ref{chirality_wilson}, for a fine $16^4$ improved
gauge configuration with $Q = 2$. The {\textsc{hyp}} action gives
results very similar to {\textsc{hisq}} and is not plotted here for
clarity.}

\label{chirality_lw}
\end{figure}

We show in fig. \ref{chirality_wilson} and \ref{chirality_lw}, for
several staggered Dirac operators, the absolute value of the first
low-lying eigenvalues, as well as the corresponding chirality, for
typical configurations with topological charge
$Q~=~2$. Fig. \ref{chirality_wilson} corresponds to a configuration
generated with the Wilson gauge action, whereas
fig. \ref{chirality_lw} is for the improved gauge action.  Due to the
exact symmetry (\ref{pairs}), we plot only half of the spectrum (in
particular, there are an equal number of near-zero modes in the other
half, and therefore only $2 Q$ of them are seen in the figures.)

We can see a strong difference between the Wilson and the improved
gauge configurations. For the one-link operator neither of them show
much of the continuum-like behaviour. As we improve the operator, the
agreement for the Wilson glue is at best qualitative, showing an
increased chirality for the low eigenmodes, and a hint of the expected
degeneracy in the spectrum. For the improved gauge glue, however, we
see the non-zero eigenvalues clearly grouping in quadruplets, and a
sharp separation between chiral and non-chiral modes. The Index
Theorem is well approximated for the {\textsc{hisq}} operator, with
the expected number of near-chiral modes.

\begin{figure}
\includegraphics[width=2.5in,clip]{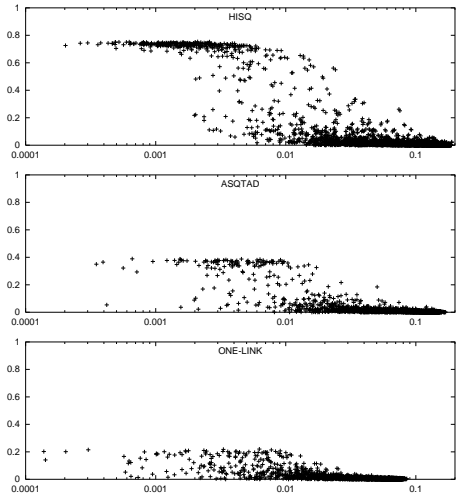}

\caption{\label{fig_scatter} A scatter plot for different staggered
quark formulations, with absolute value of the chirality on the y axis
and eigenvalue $\lambda_s$ on the x axis. The lowest 50 eigenvalues
for 147 configurations are plotted. }

\label{scatter}
\end{figure}

To give an idea of how generic this behaviour is, we show in
fig. \ref{scatter} a scatter plot of the absolute value of the
chirality vs the absolute value of the eigenvalues for a number of
improved gauge configurations. As we improve the Dirac operator, a gap
develops between the high-chirality, near-zero modes, and the
low-chirality, non-zero modes. The chirality of the near-zero modes is
remarkably constant over the different configurations, with a value of
around $0.7$.  The separation is not strict, however, and even for the
{\textsc{hisq}} operator there are configurations with low modes of
intermediate chirality. If we choose some arbitrary threshold in the
chirality to count zero modes, let's say over $.65$ in absolute value,
we can then use the Index Theorem to assign a fermionic topological
charge to the gauge field, $Q_F~=~n^+~-~n^-$. This charge coincides
with the one measured via cooling in about $90\%$ of the cases, and
therefore the ambiguity in this definition is about the same as there
is between the two cooling methods for this value of the lattice
spacing.

\begin{figure}
\includegraphics*[width=2.5in]{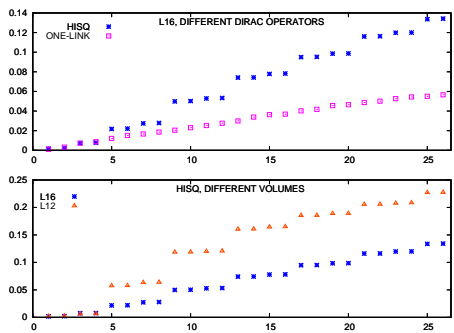}

\caption{Low lying (half) spectrum. The top panel is a comparison of
the one-link and the {\textsc{hisq}} spectrum for a configuration in the fine
$16^4$ ensemble. The bottom panel shows the {\textsc{hisq}} spectrum for the
same configuration above, and for a configuration at the same lattice
spacing and with $12^4$ volume.}
\label{spectrum}
\end{figure}

In figure \ref{spectrum} we show the effect on the spectrum of
changing the volume (keeping the lattice spacing constant) while
keeping the Dirac operator fixed, compared with changing the operator
at constant volume. As before, we need to use the improved operator
for see any degeneracy. However the lattice volume has a strong effect
too, and the degeneracy is most clear at the smaller volume. If we
were to increase the volume further, eventually the degeneracy would
not be evident any more (the spectrum will become dense in the
infinite volume limit). However, even when not obvious, it may still
be present in some form, as will be clear in our results for the
ratios of eigenvalues.

\section{RANDOM MATRIX THEORY PREDICTIONS}

Based on 
\cite{Leutwyler:1992yt},
it has been suggested that, in the $\epsilon$ regime of QCD (volume
very large, but much smaller than the pion length scale), and in each
sector of fixed topological charge $Q$, the non-zero low-lying
eigenmodes, appropriately scaled, take values from a universal
distribution, which only depends on $Q$.
\cite{Shuryak:1993pi}.
The universality class is determined by the chiral symmetries of QCD
\footnote{And also the number of dimensions, the gauge group and the
representation in which the fermions lie.}. The distributions can be
derived from any theory in the correct universality class, such as
ensembles of random matrices
\cite{Nishigaki:1998is,Damgaard:2000ah}
(for a review of other theories, see
\cite{Damgaard:2001ep}).

These predictions have been succesfully tested in lattice QCD with
Ginsparg-Wilson fermions
\cite{Edwards:1999ra,Edwards:1999zm,Damgaard:1999tk,Hasenfratz:2002rp,Bietenholz:2003mi,Giusti:2003gf}.

On the other hand, previous studies with unimproved staggered fermions
(on much coarse lattices and with Wilson action gauge fields) had
shown a very different behaviour, with the eigenvalues for any $Q$
behaving according to the theoretical prediction for the $Q = 0$
sector
\footnote{It should be noted that the method we follow here, of
grouping the eigenvalues into quartets, was not followed because this
feature of the spectrum was not evident.}
\cite{Berbenni-Bitsch:1998tx,Damgaard:1998ie,Gockeler:1998jj,Damgaard:1999bq,Damgaard:2000qt}.

\begin{figure}
\includegraphics[width=2.7in,clip]{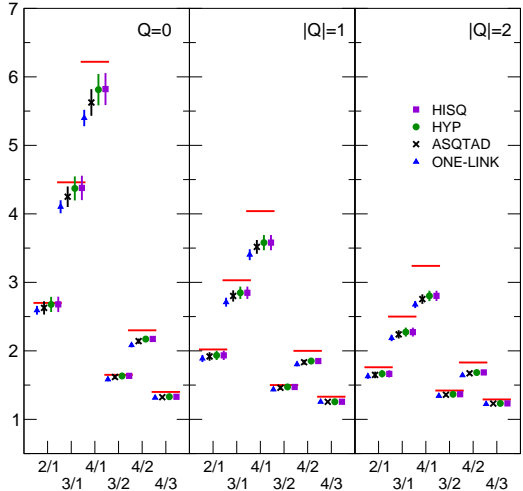}

\caption{\label{fig_mean_eig} The ratios of expectation values of
  small eigenvalues (see text for notation) compared with the
  predictions based on a universal distribution (horizontal lines) for
  topological charge sectors 0, 1 and 2.}
\label{ratios}
\end{figure}

In order to test the universality predictions, first we subtract from
the spectrum the $4Q$ lowest eigenvalues ($2Q$ on either side of
zero), which should converge to zero modes in the continuum limit,
according to the Index Theorem. We then group the remaining
eigenvalues, ordered by size, into sets of four, corresponding to the
four-fold degeneracy of the spectrum in the continuum limit. We then
average the values in each quartet, and denote the resulting averages
$\Lambda_1, \Lambda_2, \ldots$. In fig. \ref{ratios} we plot the
ratios (denoted by ``s/t'')
$\langle \Lambda_s \rangle_Q / \langle \Lambda_t \rangle_Q$
where the expectation values $\langle \cdot \rangle_Q$ are over the
sectors with gluonic topological charges $\pm Q$ only. The universal
predictions for this ratios (which are independent of any scale) are
also shown on the figure.

The first thing to notice is the clear dependence on $Q$, in stark
contrast with previous results. The ratios are systematically below
the theoretical predictions, especially the ones involving higher
eigenvalues. This would be consistent with finite volume effects as in
\cite{Giusti:2003gf}. 
There is also a small but systematic difference between the one-link
and the improved actions, with the improved results showing a better
agreement with the theoretical values. As in 
\cite{Giusti:2003gf}
we find no significant changes on the coarse lattice at the same $V$.

An important point to make here is that it is necessary to group the
eigenvalues as explained above to get sensible results. If one ignores
the near zero modes, or does not group in quartets, ratios which are
close to one or very large will result.  This is strong evidence that
the four tastes are showing up in the spectrum, even where it is not
directly evident in the spectrum itself.  

As we discussed before, there is a (small) ambiguity in the
determination of the topological charge, so that the two cooling
methods sometimes give a different answer for $Q$. One can then either
choose any one of the two methods for defining $Q$, or use in the
analysis only the configurations for which both methods agree. We have
checked that it makes no difference to within our statistical
accuracy.

\section{CONCLUSIONS AND OUTLOOK}

Improved staggered fermions are not blind to the topology, but in fact
reproduce well the predictions of the Index Theorem, and the
universality of ratios of eigenvalues as a function of topological
sector.  This means we can have confidence in using them to attack the
questions arising from the axial anomaly in QCD.

We also remark that the fact that the 4-fold taste degeneracy of
staggered quarks is becoming clear in the spectrum is encouraging for
the programme of establishing the effect of taking the fourth root of
the staggered determinant to represent one flavour of staggered sea
quarks. This programme requires an analysis in the taste basis and
progress towards this is now possible.

More extensive studies of finite volume and lattice spacing effects
and analysis of the eigenvectors are underway and will be reported
elsewhere.

Finally we would like to point out two talks on related topics in this
conference
\cite{Wenger,Wong}, 
showing similarly encouraging aspects of the staggered eigenvalue
spectrum.

\section{ACKNOWLEDGEMENTS}
    
I would like to thank the organizers of Lattice 2004 for the
opportunity of presenting this work in the plenary session. Thanks to
Ph. de Forcrand for his topological charge measurement code,
A. Hasenfratz for help in implementing the {\textsc{hyp}} operator,
and P. Lepage for many useful discussions. This work was supported by
PPARC and the EU. The eigenvalue calculations were carried out on
computer clusters at Scotgrid and the Dallas Southern Methodist
University. We thank David Martin and Kent Hornbostel for assistance.



\begin{thebibliography}{100}

\bibitem{MILC}
 MILC Collaboration: C. Aubin, C. Bernard, C. DeTar, Steven Gottlieb,
 E.B. Gregory, U.M. Heller, J.E. Hetrick, J. Osborn, R. Sugar,
 D. Toussaint
\newblock  hep-lat/0407028. 

\bibitem{Davies:2003ik}
HPQCD, C.~T.~H. Davies {\em et~al.},
\newblock Phys. Rev. Lett. {\bf 92}, 022001 (2004), [hep-lat/0304004].

\bibitem{Naik}
S. Naik, 
\newblock Nucl. Phys. B316, 238 (1989).

\bibitem{Lepage:1998vj}
G.~P. Lepage,
\newblock Phys. Rev. {\bf D59}, 074502 (1999), [hep-lat/9809157].

\bibitem{Blum:1997uf}
T.~Blum {\em et~al.},
\newblock Phys. Rev. {\bf D55}, 1133 (1997), [hep-lat/9609036].

\bibitem{Bernard:1999xx}
MILC, C.~W. Bernard {\em et~al.},
\newblock Phys. Rev. {\bf D61}, 111502 (2000), [hep-lat/9912018].

\bibitem{Orginos:1999cr}
MILC, K.~Orginos, D.~Toussaint and R.~L. Sugar,
\newblock Phys. Rev. {\bf D60}, 054503 (1999), [hep-lat/9903032].

\bibitem{Knechtli:2000ku}
A.~Hasenfratz and F.~Knechtli,
\newblock Phys. Rev. {\bf D64}, 034504 (2001), [hep-lat/0103029].

\bibitem{Follana:2003}
E.~Follana {\em et~al.},
\newblock Nucl. Phys. Proc. Suppl. {\bf 129}, 384 (2004).

\bibitem{Follana:2004}
E.~Follana {\em et~al.},
\newblock ``{F}urther improvements to staggered quarks on the lattice'',
\newblock in preparation.

\bibitem{Atiyah:1963}
M.~Atiyah and I.~Singer,
\newblock Bull. Amer. Math. Soc. {\bf 69}, 422 (1963).

\bibitem{Atiyah:1968}
M.~Atiyah and I.~Singer,
\newblock Ann. Math. {\bf 87}, 596 (1968).

\bibitem{FP}
P. Hasenfratz, V. Laliena, F. Niedermayer
\newblock Phys.Lett. B427 (1998) 125-131, [hep-lat/hep-lat/9801021].

\bibitem{Damgaard:2000ah}
P.~H. Damgaard and S.~M. Nishigaki,
\newblock Phys. Rev. {\bf D63}, 045012 (2001), [hep-th/0006111],
\newblock N.B. important correction in eprint version.

\bibitem{Damgaard:2001ep}
P.~H. Damgaard,
\newblock Nucl. Phys. Proc. Suppl. {\bf 106}, 29 (2002), [hep-lat/0110192].

\bibitem{Smit:1987fn}
J.~Smit and J.~C. Vink,
\newblock Nucl. Phys. {\bf B286}, 485 (1987).

\bibitem{Hands:1990}
S.~J. Hands and M.~Teper,
\newblock Nucl. Phys. {\bf B347}, 819 (1990).

\bibitem{Venkataraman:1998yj}
L.~Venkataraman and G.~Kilcup,
\newblock Nucl. Phys. Proc. Suppl. {\bf 63}, 826 (1998), [hep-lat/9710086].

\bibitem{Hasenfratz:2003}
A.~Hasenfratz,
\newblock ``{H}ow good is the {HYP} staggered action?'',
\newblock Talk presented at Lattice 2003 (Tsukuba, Japan).

\bibitem{Golterman:1984}
M.~Golterman and J.~Smit,
\newblock Nucl. Phys. {\bf B245}, 61 (1984).

\bibitem{Leutwyler:1992yt}
H.~Leutwyler and A.~Smilga,
\newblock Phys. Rev. {\bf D46}, 5607 (1992).

\bibitem{Shuryak:1993pi}
E.~V. Shuryak and J.~J.~M. Verbaarschot,
\newblock Nucl. Phys. {\bf A560}, 306 (1993), [hep-th/9212088].

\bibitem{Nishigaki:1998is}
S.~M. Nishigaki, P.~H. Damgaard and T.~Wettig,
\newblock Phys. Rev. {\bf D58}, 087704 (1998), [hep-th/9803007].

\bibitem{Edwards:1999ra}
R.~G. Edwards, U.~M. Heller, J.~E. Kiskis and R.~Narayanan,
\newblock Phys. Rev. Lett. {\bf 82}, 4188 (1999), [hep-th/9902117].

\bibitem{Edwards:1999zm}
R.~G. Edwards, U.~M. Heller, J.~E. Kiskis and R.~Narayanan,
\newblock Phys. Rev. {\bf D61}, 074504 (2000), [hep-lat/9910041].

\bibitem{Damgaard:1999tk}
P.~H. Damgaard, R.~G. Edwards, U.~M. Heller and R.~Narayanan,
\newblock Phys. Rev. {\bf D61}, 094503 (2000), [hep-lat/9907016].

\bibitem{Hasenfratz:2002rp}
P.~Hasenfratz, S.~Hauswirth, T.~Jorg, F.~Niedermayer and K.~Holland,
\newblock Nucl. Phys. {\bf B643}, 280 (2002), [hep-lat/0205010].

\bibitem{Bietenholz:2003mi}
W.~Bietenholz, K.~Jansen and S.~Shcheredin,
\newblock JHEP {\bf 07}, 033 (2003), [hep-lat/0306022].

\bibitem{Giusti:2003gf}
L.~Giusti, M.~Luscher, P.~Weisz and H.~Wittig,
\newblock JHEP {\bf 11}, 023 (2003), [hep-lat/0309189].

\bibitem{Berbenni-Bitsch:1998tx}
M.~E. Berbenni-Bitsch, S.~Meyer, A.~Schafer, J.~J.~M. Verbaarschot and
  T.~Wettig,
\newblock Phys. Rev. Lett. {\bf 80}, 1146 (1998), [hep-lat/9704018].

\bibitem{Damgaard:1998ie}
P.~H. Damgaard, U.~M. Heller and A.~Krasnitz,
\newblock Phys. Lett. {\bf B445}, 366 (1999), [hep-lat/9810060].

\bibitem{Gockeler:1998jj}
M.~Gockeler, H.~Hehl, P.~E.~L. Rakow, A.~Schafer and T.~Wettig,
\newblock Phys. Rev. {\bf D59}, 094503 (1999), [hep-lat/9811018].

\bibitem{Damgaard:1999bq}
P.~H. Damgaard, U.~M. Heller, R.~Niclasen and K.~Rummukainen,
\newblock Phys. Rev. {\bf D61}, 014501 (2000), [hep-lat/9907019].

\bibitem{Damgaard:2000qt}
P.~H. Damgaard, U.~M. Heller, R.~Niclasen and K.~Rummukainen,
\newblock Phys. Lett. {\bf B495}, 263 (2000), [hep-lat/0007041].

\bibitem{Wenger}
S.~Duerr, C.~Hoelbling and U.~Wenger, this conference,
\newblock hep-lat/0406027. 

\bibitem{Wong}
 K. Y. Wong, R.M. Woloshyn, this conference,
\newblock hep-lat/0407003.

\end{thebibliography}
\end{document}